\documentclass[aps,prl,twocolumn,amsmath,showpacs,superscriptaddress]{revtex4}
\usepackage[T2A]{fontenc}
\usepackage[english]{babel}
\usepackage{graphicx}
\usepackage{color}
\usepackage{bm}
\usepackage{amssymb}

\begin{document}
\title{Conditional probability calculations for \\
the nonlinear Schr\"{o}dinger equation with additive noise}

\author{I.S. Terekhov}

\email{I.S.Terekhov@inp.nsk.su}
\affiliation{Budker Institute of Nuclear Physics of SB RAS, Novosibirsk, 630090 Russia}
\affiliation{Novosibirsk State University, Novosibirsk, 630090, Russia}

\author{S.S. Vergeles}

\affiliation{Landau Institute for Theoretical Physics, Moscow, 119334, Russia}

\author{S.K. Turitsyn}

\affiliation{Aston Institute of Photonic Technologies, Aston University, Aston Triangle, Birmingham B4 7ET, UK }

\affiliation{Novosibirsk State University, Novosibirsk, 630090, Russia}

\begin{abstract}
The method for computation of conditional probability density function for the nonlinear Schr\"{o}dinger equation with additive noise is developed.
We present in a constructive form the conditional probability density function in the limit of a small noise and analytically derive it in a weakly nonlinear case.
The general theory results are illustrated using fibre-optic communications as a particular, albeit practically very important, example.
\end{abstract}

\pacs{05.10.Gg, 89.70.-a,  02.70.-c,02.70.Rr,05.90.+m}
\maketitle

The nonlinear Schr\"{o}dinger equation (NLSE) is one of the most general and fundamental models of nonlinear science, with a vast number of applications ranging from hydrodynamics, plasma physics and bio-physics to modern high-speed fibre-optic communications (see e.g. \cite{N,KA,ZMNP,AS,Sulem,Kuz,IMMS,zs72,TBF,Narimanov2002,Essiambre2008} and references therein). The NLSE, in particular, describes in the main order high-frequency wave propagation in media with nonlinear and dispersive effects,  making it a very generic mathematical model.  The NLSE is also of special interest, because it presents an example of an integrable nonlinear system with an infinite number of degrees of freedom \cite{zs72}. Here, without loss of generality, we discuss NLSE in the practically important context of  optical communications, however, the obtained mathematical results are very general and may be applied in a wide range of physics problems. We present a method for computing  a conditional probability density function (PDF) for the NLSE with  additive white Gaussian noise. We apply the developed method for derivation of an analytical expression for PDF in the practically important limit of weak nonlinearity.

The article is organized as follows. First, we show that instead of direct massive statistical numerical modeling (e.g. using Monte Carlo
methods) of the NLSE with different realizations of the noise one can calculate numerically a path-integral where
integration over different noise realizations is done analytically. This is an important and non-trivial step change simplifying the overall numerical modeling of the probability density function for the equation of high practical importance (in the optical communication context)
and of wide applicability across many areas of physics. Then we demonstrate that in the case when signal-to-noise power ratio is large, the main contribution to the path-integral gives "classical trajectory".  We found the equation for the "classical trajectory", and the solution of this equation gives main contribution to the conditional probability density function. Then we demonstrate the application of our method to calculation of conditional probability density function for specific example. We would like to stress that the proposed methodology of calculation of the general conditional probability  is applied to arbitrary field at destination (received signal) and as such, cannot be practically obtained through direct Monte Carlo modelling of the NLSE for different noise realisations.

Consider the nonlinear Schr\"{o}dinger equation for field $\psi_\omega (z)$ with additive noise $\eta_\omega (z)$ presented here in the frequency domain:
\begin{eqnarray}\label{startingCannelEq}
\partial_z \psi_\omega (z)&=&i\frac{\beta_2}{2}\omega^2\psi_\omega (z)+ \eta_\omega (z)+\nonumber\\
&&i\gamma\int\frac{d\omega_1d\omega_2}{(2\pi)^2}
\psi_{\omega_1} (z) \psi_{\omega_2} (z) \bar{\psi}_{\omega_3} (z)\,.
\end{eqnarray}
Here and in what follows $\omega_3=\omega_1+\omega_2-\omega$. In the optical-fibre applications context, $\beta_2$ is the group velocity dispersion parameter, $\gamma$ is the Kerr nonlinearity coefficient, bar means complex conjugation, $\eta_\omega (z)$ is an additive complex white noise (resulting in fibre communication applications from optical amplification) with zero mean and correlation function \cite{IMMS,Essiambre2008}:
$
\langle \eta_\omega (z)\bar{\eta}_{\omega^\prime} (z^\prime)\rangle_{\eta} = 2\pi Q \delta(z-z^\prime)\delta(\omega-\omega^\prime)\,.
$
Using the Martin-Siggia-Rose formalism \cite{Zinn-Justin,Lebedev,SuplMat}, we can formally present the conditional probability density $P[Y(\omega)|X(\omega)]$ to have $\psi_\omega(L)=Y(\omega)$ if $\psi_\omega(0)=X(\omega)$ in the form of the Feynman path integral, corresponding to the Eq. (\ref{startingCannelEq}):
\begin{eqnarray}\label{probabilityInitial}
P[Y(\omega)|X(\omega)]=\int\limits_{\psi_\omega (0)=X(\omega)}^{\psi_\omega (L)=Y(\omega)}{\cal D}\psi \exp\left\{-\frac{S[\psi]}{Q}\right\}\,,
\end{eqnarray}
where the action $S[\psi]$ reads:
\begin{eqnarray}
&&S[\psi]=\int_0^L dz\int\frac{d\omega}{2\pi}\left|{\cal L}^{(0)}[\psi]-V[\psi] \right|^2\,,\\
&&{\cal L}^{(0)}[\psi]=\partial_z \psi_\omega (z)-i\frac{\beta_2}{2}\omega^2\psi_\omega (z)\,,\\
&&V[\psi]=i\gamma\int \frac{d\omega_1d\omega_2}{(2\pi)^2}
 \psi_{\omega_1} (z) \psi_{\omega_2} (z) \bar{\psi}_{\omega_3} (z)\,,
\end{eqnarray}
with the measure ${\cal D}\psi$  defined as:
\begin{eqnarray}
{\cal D}\psi=\underset{M\to\infty}{\lim_{\delta\to 0
}}\underset{N\to\infty}{\lim_{\Delta\to 0 }} \left(\frac{\delta}{\Delta\pi Q}\right)^{NM} \prod_{j=1}^M \prod_{i=1}^N d \psi_{i,j}\,,
\end{eqnarray}
here $d\psi_{i,j}=dRe\{\psi_{i,j}\}d Im \{\psi_{i,j}\}$, the first and second indexes in the $\psi_{i,j}$ enumerate $z$ and $\omega$ coordinates, respectively,  see below. Since the action $S[\psi]$ is not a quadratic form in the functions $\psi_\omega (z)$ and $\bar{\psi}_\omega (z)$, the path integral  Eq. (\ref{probabilityInitial}), in a general case, can not be calculated analytically. The examples when the path integral for PDF can be derived analytically include the case of the zero dispersion, $\beta_2=0$ \cite{Turitsyn2003}, and a specific constraint on the initial field to be a soliton \cite{FKLT,DTY}. However, the presentation of the PDF in the form of the integral Eq. (\ref{probabilityInitial}) is convenient for numerical computation and for using a perturbation theory. For the purpose of numerical calculations of the PDF, Eq. (\ref{probabilityInitial}) should be presented in a discrete form. Taking into account the causality principle, which in our case means that the function $\psi_\omega (z)$ is affected only by the dynamics of $\psi_\omega (z^\prime)$ in the preceding  points of evolution at $ z^\prime<z $, we obtain:
\begin{widetext}
\begin{eqnarray}\label{probabilityDiscretForm}
P[Y(\omega)|X(\omega)] &=& \underset{M\to\infty}{\lim_{\delta\to 0
}}\underset{N\to\infty}{\lim_{\Delta\to 0 }}\tilde{\Lambda} \int\left[\prod_{j=1}^M \prod_{i=1}^{N-1} d \psi_{i,j}\right]
 \exp\left\{-\frac{\Delta\delta}{Q}\sum_{i=1}^{N}\sum_{j=1}^{M} \left|\frac{\delta\psi_{i,j}}{\Delta}-i\frac{\beta_2}{2}\omega_j^2\psi_{i-1,j} -V_{i-1,j}\right|^2\right\}\,.
\end{eqnarray}
\end{widetext}
Here $\tilde{\Lambda}=\left(\Delta\pi Q/\delta\right)^{-NM}$, $\psi_{i,j}=\psi_{\omega_j}(z_i)$,  $\delta\psi_{i,j}=\psi_{i,j}-\psi_{i-1,j}$, $z_i= i\Delta$, $i=0,1,...,N$, $z_N=L$, $\omega_j=\Omega_{min}+2\pi(j-1)\delta$, $j=1,2,...,M$, $\omega_M=\Omega_{max}$, $W=\Omega_{max}-\Omega_{min}$.  In Eq. (\ref{probabilityDiscretForm})  we took into account the boundary conditions: $\psi_{0,j}=X(\omega_j)=X_j$, $\psi_{N,j}=Y(\omega_j)=Y_j$. The conditional probability satisfies the standard condition $\int{\cal D}Y P[Y(\omega)|X(\omega)]=1\,$,
where ${\cal D}Y=\prod\limits_{j=1}^M dY_j$, (see for details  Eq. (13) in \cite{SuplMat}). Equation (\ref{probabilityDiscretForm}) is the first important result of our work. The $2M(N-1)$-fold integral can be calculated numerically with the required accuracy using Monte-Carlo methods, see e.g. Ref.\cite{Pierro2001}. Therefore Eq. (\ref{probabilityDiscretForm}) provides a constructive way to compute PDF for the NLSE in most general cases.

Moreover, the presentation (\ref{probabilityDiscretForm}) allows us to develop the perturbation theory using small nonlinearity parameter and derive an analytical expression for the conditional probability. To do so, let us introduce two dimensionless parameters, $\tilde{\gamma}=\gamma P_{ave} L$ and $\epsilon=QLW/(2\pi P_{ave})=1/\mathrm{SNR}$, where $P_{ave}=T_{total}^{-1}\int \dfrac{ d\omega}{2\pi} |X(\omega)|^2$ is the average power of the signal, $T_{total}$ is the full time interval of a signal pattern, $W/2\pi$ is the noise/channel bandwidth. The dimensionless parameter $\epsilon$ is nothing more, but the inverse power signal-to-noise-ratio (SNR). The dimensionless parameter $\tilde{\gamma}$  describes the effective nonlinearity. Later we impose that $\tilde{\gamma}\ll 1$ and develop the perturbation theory in the parameter $\tilde{\gamma}$ for different values of $\epsilon$. In the case $\tilde{\gamma}/\epsilon\ll 1$ we can expand exponential function in Eq. (\ref{probabilityInitial}). When the parameter $\tilde{\gamma}\ll 1$ and $\tilde{\gamma}/\epsilon\sim 1$, or even $\tilde{\gamma}/\epsilon\gg 1$ we use a method similar to that developed in quantum mechanics for finding the classical trajectory of the particle.

Let us start the consideration from the case $\tilde{\gamma}/\epsilon\ll 1$. Using standard methods of quantum field theory, see \cite{Zinn-Justin,IZ}, we expand the exponent in Eq. (\ref{probabilityInitial}) at small $\gamma$. After that the function $P[Y(\omega)|X(\omega)]$ can be represented as a series in $\gamma$:
\begin{eqnarray}\label{series}
 P[Y(\omega)|X(\omega)]= \sum_{n=0}^\infty \frac{\gamma^n}{n!} P^{(n)}_{(\gamma)}[Y(\omega)|X(\omega)]\,,
\end{eqnarray}
In the zero order in $\gamma$ we  obtain an effective Gaussian channel approximation, see \cite{SuplMat}:
\begin{eqnarray}\label{probabilityFormZeroFinal}
P^{(0)}_{(\gamma)}[Y(\omega)|X(\omega)] = \Lambda\exp\left\{-\frac{1}{QL}\int\frac{d\omega}{2\pi}\left|B(\omega)\right|^2\right\}\,,
\end{eqnarray}
where $\Lambda$ is the normalization constant, $\Lambda=P^{(0)}_{(\gamma)}[0|0]=(\pi Q L/\delta)^{-M}$, $B(\omega)=Y(\omega)e^{-i\beta_2\omega^2L/2}-X(\omega)$. The function $B(\omega)$ proportiional to difference  of the  $Y(\omega)$ and the solution $\psi^{(0)}_\omega(L)$ of the Eq. (\ref{startingCannelEq}) with $\gamma=0$, $\eta=0$, and boundary condition $\psi_\omega(0)=X(\omega)$, therefore $P^{(0)}_{(\gamma)}[Y(\omega)|X(\omega)]$ is the Gauss  distribution of functions around  $\psi^{(0)}_\omega(L)$ in functional space. It is easy to see that $P^{(0)}_{(\gamma)}[Y(\omega)|X(\omega)]$ is normalized as $\int{\cal D}Y P^{(0)}_{(\gamma)}[Y(\omega)|X(\omega)]=1$. This means that all corrections in $\gamma$ are normalized here to satisfy condition $\int{\cal D}YP^{(n\ne0)}_{(\gamma)}[Y(\omega)|X(\omega)]=0$.

The higher-order corrections in $\gamma$ can be calculated in any order from Eq. (\ref{probabilityDiscretForm}) (see Supplementary Materials for details).
As an example, we write down here the first order correction $P^{(1)}_{(\gamma)}[Y(\omega)|X(\omega)]$. Using the procedure described in the Supplementary Materials  we obtain:
\begin{eqnarray}\label{smallGammafirstOrderFinal}
&&\!\!\!\!\!\!\!\!\!P^{(1)}_{(\gamma)}[Y(\omega)|X(\omega)]= P^{(0)}_{(\gamma)}[Y(\omega)|X(\omega)]\times \nonumber\\
&&\!\!\!\!\!\!\!\!\! Im\Bigg\{\frac{ W L}{3\pi}\int\frac{d\omega}{2\pi}e^{-i\beta_2\omega^2 L/2}Y(\omega)\bar{X}(\omega)+G\Bigg\}\,,\\
&&\!\!\!\!\!\!\!\!\!G=\frac{2}{Q}\!\!\int\limits_0^L\!\!\frac{dz}{L}\!\!\!\int\!\frac{d\omega d\omega_1 d\omega_2}{(2\pi)^3}e^{\mu z}B(\omega)
\bar{\lambda}_{\omega_1}(z)\bar{\lambda}_{\omega_2}(z)\lambda_{\omega_3}(z) \,, \nonumber\\
&&\!\!\!\!\!\!\!\!\!\lambda_\omega(z)=X(\omega)+zB(\omega)/L.
\end{eqnarray}
Here $\mu=i\beta_2 (\omega-\omega_1)(\omega-\omega_2)L$. The result contains two different terms, the first one is proportional to the bandwidth $W$ and does not involve parameter $Q$, the other one (function $G$) has $Q$ in the denominator. Therefore in this limit (small $\gamma$) of the perturbation theory the noise parameter $Q$ is assumed to be not too small. In physical terms this is the limit of a weakly nonlinear and highly noisy system.

In the different and practically important limit of  small $\epsilon$ or $Q$ (large SNR) the conditional probability can be computed using a method similar to that one used to calculate classical trajectory in quantum mechanics \cite{Feinman}. In simple terms, this corresponds to finding the solution without the noise term and making a functional expansion around this solution due to the small noise (high signal-to-noise ratio). In the case under consideration we can use the Laplace's method, see e.g. \cite{LS}. The main contribution to path integral in Eq. (\ref{probabilityInitial}) gives the region around the trajectory where the action $S[\psi]$ reaches the minimum. Let $S$ approach the minimum at the trajectory $\Psi_\omega(z)$,  Eq. (\ref{probabilityInitial}) can be rewritten in the following form
\begin{eqnarray}\label{QuasiclassProbabInitial}
&&P[Y(\omega)|X(\omega)] = e^{-S[\Psi_\omega (z)]/Q}\times\nonumber\\
&&\int\limits_{\tilde{\psi}_\omega (0)=0}^{\tilde{\psi}_\omega (L)=0}{\cal D}\tilde{\psi} e^{-(S[\Psi_\omega (z)+\tilde{\psi}_\omega (z)]-S[\Psi_\omega (z)])/Q}\,,
\end{eqnarray}
the explicit form of the path integral is shown in \cite{SuplMat}. Thus, the problem of calculation of the conditional probability reduces to finding function $\Psi_\omega(z)$ and calculation of the path integral with zero boundary conditions. We would like to emphasize once more the important difference of the proposed approach and direct Monte Carlo modelling of the NLSE with different realisations of noise. In the  path-integral method we can constructively compute probability density function for arbitrary received signal $Y(\omega)$, even for PDF with very rare events while in the direct modelling of the NLSE it might be practically impossible to find trajectories with low probability, that still can be important for the system performance.  Now  we calculate the conditional probability in the leading orders in $Q$.

To calculate the path integral we expand the expression in the exponent in path integral at small $\tilde{\psi}$  to the series in $\tilde{\psi}$. Since $S$ reaches the minimum at  $\Psi_\omega(z)$, the series starts from the second order in $\tilde{\psi}$. To calculate the path integral in the leading order in $Q$ we keep terms only in the main order in $\tilde{\psi}$ in the series. Then we calculate the integral using the perturbation theory in $\gamma$ developed in \cite{SuplMat}, and obtain the result in the leading and next to the leading order in $\gamma$:
\begin{widetext}
\begin{eqnarray}\label{functional denominator}
P[Y(\omega)|X(\omega)]\approx \Lambda e^{-S[\Psi_\omega (z)]/Q}\left(1+\frac{2\gamma W}{\pi}Im\left\{\int\limits_0^L dz\frac{z(L-z)}{L}\int\frac{d\omega}{2\pi}{\cal L}^{(0)}\left[\Psi_\omega (z)\right]\bar{\Psi}_\omega (z)\right\}\right)\,.
\end{eqnarray}
It is seen that in order  to calculate $P[Y(\omega)|X(\omega)]$ we first have to determine function $\Psi_\omega(z)$ ("classical trajectory"). The action approaches the minimum at  $\Psi_\omega(z)$, therefore, $\delta S[\Psi]=0$, where $\delta S$ is a variation of $S$. This last equation leads to the following equation for $\Psi_\omega(z)$
(analogue of a classical trajectory):
\begin{eqnarray}\label{classicalTrajectoryEq}
&&\left(\partial_z-i\frac{\beta_2\omega^2}{2}\right)^2\Psi_\omega (z) - i\gamma\int\frac{d\omega_1d\omega_2}{(2\pi)^2}
\bigg\{4 \Psi_{\omega_2} (z) \bar{\Psi}_{\omega_3} (z)\left[\left(\partial_z-i\frac{\beta_2\omega_1^2}{2}\right)\Psi_{\omega_1} (z)\right]-\frac{\mu}{L}\Psi_{\omega_1} (z) \Psi_{\omega_2} (z) \bar{\Psi}_{\omega_3} (z)\bigg\}-\nonumber\\
&&3\gamma^2 \int\frac{d\omega_1d\omega_2d\omega_4 d\omega_5d\omega_6}{(2\pi)^4} \delta(\omega_1+\omega_2+\omega_4-\omega_5-\omega_6-\omega) \Psi_{\omega_1} (z) \Psi_{\omega_2} (z) \Psi_{\omega_4} (z)\bar{\Psi}_{\omega_5} (z)\bar{\Psi}_{\omega_6} (z)=0\,,
\end{eqnarray}
\end{widetext}
with the boundary conditions: $\Psi_\omega (0)=X(\omega)\,,\quad \Psi_\omega (L)=Y(\omega)$. Equation (\ref{classicalTrajectoryEq}) can be written in time domain, see \cite{SuplMat}.

Calculating action $S[\Psi_{\omega}(z)]$ up to the first order in $\gamma$ yields (see for details Supplementary Materials):
\begin{eqnarray}\label{smallQfirstOrderFinal}
&&\!\!\!\!\!\!\!\!P[Y(\omega)|X(\omega)]\approx P^{(0)}_{(\gamma)}[Y(\omega)|X(\omega)]e^{\gamma Im\{G\}}
\nonumber\\
&&\!\!\!\!\!\!\!\!\left(1+\frac{\gamma W L}{3\pi}Im\left\{\int\frac{d\omega}{2\pi}e^{-i\beta_2\omega^2 L/2}Y(\omega)\bar{X}(\omega)\right\}\right)\,.
\end{eqnarray}
Note that the exponent $e^{\gamma Im\{G\}}$ can not be expand at small $\gamma$ in general case, since we have assumed here that the parameter $\epsilon\ll 1$.  However, when $\tilde{\gamma}/\epsilon\ll1$ the result (\ref{smallQfirstOrderFinal}) coincides with $P^{(0)}_{(\gamma)}[Y(\omega)|X(\omega)]+\gamma P^{(1)}_{(\gamma)}[Y(\omega)|X(\omega)]$, as it should be. The Eq. (\ref{smallQfirstOrderFinal}) is the NLSE channel PDF in the limit of small $Q$
in the leading order in $\gamma$.

Now  we illustrate the application of the derived general PDF (valid for arbitrary input  $X$), considering some particular choices of initial signal.
Let the signal in the time domain  have the form:
\begin{eqnarray}
X(t)=\sum_{k=-N}^{N}c_k F(t-k T)\,,\quad F(t)=\alpha\, e^{-t^2/2\tau^2}\,,
\end{eqnarray}
where $N\gg1$ is the number of pulses in the information pattern, $c_k=e^{\phi_k}$, where $\phi_k$ is the value, which is randomly chosen from $\{0,i\pi/2,i\pi,-i\pi/2\}$, $F(t)$ is the waveform of the carrier pulse, $T$ is the time interval between pulses (baud rate), $\tau$ is the parameter related to the pulse width, we assume here that $\tau\ll T$. The constant $\alpha$ defines the signal average power $P_{ave}=T^{-1}\int_{-\infty}^{\infty}F^2(t)dt=\alpha^2\tau\sqrt{\pi}/T$. In the frequency domain the initial signal is presented as:
\begin{eqnarray}\label{initialSignalOmega}
X(\omega)=\sqrt{2\pi}\alpha\tau e^{-\omega^2\tau^2/2}\sum_{k=-N}^{N}c_k e^{i\omega kT}\,.
\end{eqnarray}
Consider PDF distributions of $c_k$ assuming that the received signal $Y(\omega)$ can be approximated as
\begin{eqnarray}\label{finallSignalOmega}
Y(\omega)&=&\biggl\{X(\omega) + \sqrt{2\pi}\alpha\tau e^{-\omega^2\tau^2/2}\sum_{k=-N}^{N}\rho_k e^{i\phi_k}e^{i\omega k T}+\nonumber\\
&&i\gamma L\phi_{nl}(X(\omega))\biggr\}e^{i\beta_2\omega^2L/2}\,,
\end{eqnarray}
where the average nonlinear phase shift (rotation of the phase, same to all pulses) is:
\begin{eqnarray}
\!\!\phi_{nl}(X(\omega))=\int\!\frac{d\omega_1 d\omega_2}{(2\pi)^2}X(\omega_1)X(\omega_2)\bar{X}(\omega_3)\frac{1-e^{-\mu}}{\mu}.
\end{eqnarray}
This choice of the $Y(\omega)$  implies that all coefficient $c_k$ are changed  to $\tilde{c}_k=c_k+\rho_k e^{i\tilde{\phi}_k}$, which corresponds to corruption of the signal constellation points by noise and (weak) nonlinear effects. For the sake of clarity in this methodological Letter  we imply that pulses do not broadened large ($|\beta L/\tau|\ll T$). Then  we can use property (46) of \cite{SuplMat} for conditional probability. The  substitution of Eq. (\ref{finallSignalOmega}) and Eq. (\ref{initialSignalOmega}) to Eq. (\ref{smallQfirstOrderFinal})  yields the following conditional probability
\begin{eqnarray}\label{smallQfirstOrderFinalNumRes}
&&P[Y(\omega)|X(\omega)]\approx \prod_{k=-N}^N P_k\,,\\
&&P_k=\Lambda^{1/(2N+1)}\exp\left\{-\frac{P_{ave} T}{QL}\rho_k^2 \right\}\times\nonumber\\
&&\left(1+\frac{\gamma W T L P_{ave}}{3\pi}\rho_k\sin(\tilde{\phi}_k-\phi_k)\right)\,.
\end{eqnarray}
One can see that the conditional probability of the whole signal pattern is the product of conditional probabilities for each pulse as it should be for non-interfering signals. 
Of course, the general PDFs derived above do include pulse-to-pulse interference that can be accounted for perturbatively. Since $\tilde{\gamma}\ll 1$, $P_k$    is the
slightly deformed Gaussian distribution. Of course, the result (\ref{smallQfirstOrderFinalNumRes}) is formally written with excessive accuracy and should be used only in the first order of the parameter $\tilde{\gamma}\ll 1$: 
\begin{eqnarray}
&&P[Y(\omega)|X(\omega)]\approx\Lambda\exp\left\{-\frac{P_{ave} T}{QL}\sum_{k=-N}^{N}\rho_k^2 \right\}\times\nonumber\\
&&\left(1+\frac{\gamma W T L P_{ave}}{3\pi}\sum_{k=-N}^{N}\rho_k\sin(\tilde{\phi}_k-\phi_k)\right)\,.
\end{eqnarray}
Note that we already took into account overall phase shift $\phi_{nl}(X(\omega))$ in $Y(\omega)$, see Eq. (\ref{finallSignalOmega}). 
We would like to stress that Eq. (\ref{smallQfirstOrderFinalNumRes}), of course, is just a particular example of using the general formulae for the NLSE PDF derived above. In the general case, one can use either the PDF (Eq. (\ref{probabilityDiscretForm})) for numerical analysis with arbitrary input signal  or expressions Eq. (\ref{smallGammafirstOrderFinal}) and Eq. (\ref{smallQfirstOrderFinal}) for simplified numerical or analytical  analysis in practically important limits.

In conclusion,  we have introduced a constructive method for numerical computation of the conditional probability for the nonlinear Schr\"{o}dinger equation through multidimensional integrals. We have developed an analytical method for conditional probability calculation for nonlinear noisy fiber optic communication channels in the case of weak nonlinearity and arbitrary parameter $\epsilon=1/\mathrm{SNR}$ which is the inverse signal-to-noise power ratio. In the limit $\epsilon\sim 1$ we derived an equation for calculating of the conditional probability using the perturbation theory in $\gamma$. In the limit $\epsilon\ll 1$ we have derived the classical trajectory and developed a method similar to finding the quantum corrections to the classical trajectory in quantum mechanics. The  path-integral method allows one to constructively compute PDFs for any received signal $Y(\omega)$, even corresponding to very rare events, while in the direct modelling of the NLSE it might be practically impossible to find trajectories with such low probability. We believe that our results might find various applications ranging from statistical physics \cite{N1,N2,N3} to high capacity optical communications \cite{Essiambre2008,N4}. The approach provides a platform for optimization over initial signal distributions that is of critical importance for computation of the Shannon capacity of communication channels.

{\it Acknowledgements.}  We acknowledge the financial support of  the  Engineering and Physical Sciences Research Council (project UNLOC) and the grant of the Ministry of Education and Science of the Russian Federation (agreement No. 14.B25.31.0003).

\end{document}